# In search of high performance anode materials for Mg batteries: computational studies of Mg in Ge, Si, and Sn

*Oleksandr I. Malyi[1], Teck L. Tan[2], and Sergei Manzhos[1]\**

[1]Department of Mechanical Engineering, National University of Singapore, Block EA #07-08, 9 Engineering Drive 1, Singapore 117576, Singapore

[2]Institute of High Performance Computing, 1 Fusionopolis Way, #16-16 Connexis, Singapore 138632, Singapore

*\*E-mail:* mpemanzh@nus.edu.sg*; Fax: +65 6779 1459; Tel: +65 6516 4605*

ABSTRACT. We present ab initio studies of structures, energetics, and diffusion properties of Mg in Si, Ge, and Sn diamond structures to evaluate their potential as insertion type anode materials for Mg batteries. We show that Si could provide the highest specific capacities (3817 mAh g$^{-1}$) and the lowest average insertion voltage (~0.15 eV vs. Mg) for Mg storage. Nevertheless, due to its significant percent lattice expansion (~216%) and slow Mg diffusion, Sn and Ge are more attractive; both anodes have lower lattice expansions (~120 % and ~178 %, respectively) and diffusion barriers (~0.50 and ~0.70 eV, respectively, for single-Mg diffusion) than Si. We show that Mg-Mg interactions at different stages of charging can decrease significantly the diffusion barrier compared to the single atom diffusion, by up to 0.55 eV.

1. **Introduction**

Increased needs in energy storage and limited access to Li resources lead to a shift of interest to non-Li ion batteries. Metals such as Na [1-3], Mg [4, 5], Al [6] are attractive for energy storage applications because of their low cost and availability. While Na batteries are attractive and in fact used [1] for bulk storage, they are not very suitable for applications where high energy density is required, such as all-electric vehicles and portable electronics, due to the low rate capability and specific capacity of Na storage electrodes [2, 3]. The research into the potential of batteries using other metals such as Ca [7] or Al [6] is still in its infancy. Mg batteries, on the other hand, are currently emerging as a viable next generation rechargeable battery technology, so much so that commercialization is believed to be achievable with several companies working on it [4, 5, 8]. The key advantages of Mg include: (i) it results in the storage of up to 2 electrons per Mg atom vs. one for Li and Na, resulting in a higher theoretical volumetric energy density and a specific capacity comparable to those of Li in spite of being heavier than both Li and Na; (ii) metallic Mg is cheaper than metallic Li; (iii) the atomic radius of Mg is smaller than that of Na and is comparable to Li. A smaller size is expected to result in higher diffusion rates and therefore a better battery rate capability, and (iv) Mg does not have a severe dendrite formation problem that plagues metallic Li and Na [9, 10]. In fact, most works on Mg batteries [8, 11, 12] used metallic Mg anodes and focused on the development of cathode materials. The cycling ability of Mg batteries, however, remained poor. This has in particular to do with reactions of the anode with electrolyte species resulting in the formation of a blocking layer [8]. Electrode-electrolyte interactions also limit the achievable battery voltage. *High voltage Mg batteries can be developed if the metallic anode is replaced with an insertion anode* [4, 5]. Simply put, the success of Mg batteries may depend on successful design of Mg anodes. Insertion type anode materials have been extensively researched for Li batteries. Specifically, group IV elements (Ge, Si, and Sn) have been shown to result in very high Li specific capacities (up to 4400 mAh $g^{-1}$ (Si)). The possibilities of insertion anodes in Mg batteries are still largely unstudied. There have been recent and encouraging experimentations with insertion anodes. *Singh et al.* [4, 5] have reported investigations Mg batteries with Sb, Sn, and Bi anodes. Theoretical studies of Mg insertion anodes are, however, lacking but are critical to guide the design of new electrode materials. The performance of the anode material is heavily dependent on the type of metal atom: qualitative differences in insertion structures and in voltage profiles have been reported [13]. Differences in diffusion properties can potentially result in drastic differences in battery rate capabilities for different metal ions. In this study, we analyze the potential of Ge, Si, and Sn for bulk storage and the high energy density applications. Based on density functional theory (DFT) calculations (see Supporting Information), we study structures, energetics, and diffusion properties of Mg in Si, Ge, and Sn, including the effects on the diffusion barrier of Mg-Mg interactions and of high anode charge states – alloys $XMg_2$, where X= Ge, Si, and Sn. Although β-X structures can be more stable under some conditions, here, we focus on diamond like structures (α-X) as potential anode materials.

2. **Results and discussion**

It has been reported that at high dopant concentrations, Mg interacts with X forming $XMg_2$ (Fm3m space group) structures and that for a wide range of Mg concentrations, X coexists with $XMg_2$ structures [14]. Among the considered materials, Si can provide the highest specific capacity (see Table 1). For instance, the theoretical specific capacity for Si is 2906 mAh g$^{-1}$ and 565 mAh g$^{-1}$ larger than those of Sn and Ge, respectively. Taking into account that Li and Mg have one and two valence electrons, respectively, and using the methodology proposed by *Ceder et al.* [15], the average insertion voltage (*v*) vs. metallic Mg for the prospective anode materials is predicted using Eq.1:

$$V = -\frac{E(XMg_2) - E(X) - 2E(Mg)}{4e}, \qquad (1)$$

where *E* is the total energy of the reference state (see Table S1) as calculated using DFT, *e* is the absolute value of the electron charge. The lowest average voltage (0.151 V) is found for $SiMg_2$ (see Table 1), while for $GeMg_2$, the largest voltage (0.241 V) is predicted. Although both the specific capacity and the average insertion voltage are important for the performance of metal ion batteries, insertion type anode can also have a significant expansion during charging/discharging process. Anode materials having low average insertion voltages and high specific capacities but huge lattice expansions may not be appropriate for specific battery applications. This is because a significant anode deformation changes its mechanical stability [16] and consequently can adversely affect the battery's cyclability. This problem is well known for the Si anode in Li batteries and its resolution required transition to nanostructure materials [17], which also results in capacities much lower than theoretical [18-20] as well as in an increased cost. Therefore, starting with a material having a smaller expansion as well as a high capacity and suitable voltages would clearly be advantageous.

To evaluate prospective anode materials, *Obrovac et al.* [13] predicted not only voltages and specific capacities but also volumetric energy densities (Wh/cc) and the extent of the materials' expansions. Using the same methodology, the volumetric energy density ($\tilde{U}_f$) of a charged negative electrode alloy material can be calculated according to Eq. 2:

$$\tilde{U}_f = \frac{FV_{avg}}{v}\left(\frac{\xi_f}{1+\xi_f}\right), \qquad (2)$$

where $F$ is Faraday's constant (26.802 Ah mol$^{-1}$); $\xi_f$ is the final percent volume expansion (in fully charged anode material); $V_{avg}$ is the average voltage of the alloy vs. a cathode; $v$ is the volume occupied by Mg. Since in the recent studies, volumetric energy densities were predicted vs. a 3.75 V cathode, we also used that value for the reference state of cathode. Predicted volumetric energy densities show much smaller differences between the three anode materials than differences of specific capacities and average voltages. Si can provide the largest volumetric energy density (9.929 Wh cc$^{-1}$) at ~216% volume expansion; while Ge can provide 9.487 Wh cc$^{-1}$ at ~178%, as shown in Fig. 1. Figure 1 has practical significance for the design of anodes. For example, suppose we are able to design an X-based anode that withstands a 100% volume expansion, then the volumetric energy density can be predicted according to the figure. In this light, it is important to consider volumetric energy densities for the prospective anode materials at the same volume expansion. For a 100% volume expansion, Sn and Ge can provide almost the same volumetric energy densities (7.413 Wh cc$^{-1}$ and 7.382 Wh cc$^{-1}$, respectively) which are ~0.152 Wh cc$^{-1}$ larger than that for Si (7.231 Wh cc$^{-1}$). It should be noted that recently *Obrovac et al.* [21] predicted volumetric energy densities for β-Sn. They found that Si has the volumetric energy density larger by ~1.377 Wh cc$^{-1}$ compared to that of β-Sn. The difference between our and that prediction is explained by a different Mg volume occupied in α-Sn (6.42 mL mol$^{-1}$ charge$^{-1}$) and β-Sn (7.57 mL mol$^{-1}$ charge$^{-1}$) [21]. Taking into account that α-Sn/β-Sn phase transition takes place at ~287 K and can be changed by doping [22], environmental conditions [23, 24] etc., it becomes clear that the use of stabilized α-Sn can provide a much better performance compared to that of β-Sn.

Taking into account that all three materials can take up to 2 Mg atoms per host atom, for a realistic comparison of rate capabilities of different anode materials, it is also critical to investigate the diffusion behavior of Mg atoms at different concentration. For low concentrations, predicted defect formation energies (see Table S2) suggest that Mg atoms in all three prospective anode materials act as interstitial defects occupying tetragonal (T) sites (see Fig. 2(a)). This is similar to the behavior of Li [25] and Na [26] atoms in the Si matrix. Addition of a second Mg atom leads to the Mg-Mg interaction changing the stability of Mg defects. Predicted formation energies of Mg-Mg defects suggest that Mg atoms do not tend to cluster (see Fig. 3). At low dopant concentrations, the insertion process is associated with migration of Mg atoms from one T site to another. Our calculations suggest that for all considered structures, the Mg atom migrates between two T sites via a hexagonal (Hex) site (see Fig. 2(b)). Despite this similarity, Mg diffusion barriers are very different for the three systems (see Fig. 4(a)). The lowest Mg migration barrier (0.497 eV) is predicted for Sn and it is 0.503 eV smaller than that for Si. This observation suggests that the initial stage of Mg insertion is significantly faster for Sn compared to the other materials. As a second Mg atom comes into the matrix, Mg-Mg interactions change Mg diffusion behavior. Specifically, when two Mg atoms are close to each other, the predicted migration barriers are smaller compared to those for single Mg atoms (see Fig. 4(b)). This is caused by both a local expansion of the matrix and a destabilizing effect of Mg-Mg interactions (see Fig. 3). Indeed, a local expansion caused by deformation or additional atoms can reduce the migration barrier [26]. Hence, Mg-Mg interaction can increase Mg diffusivity for some range of doping concentrations. Nevertheless, due to a large energy cost of clustering, it is expected to be less likely compared to both Li and Na [26] at the concentrations considered here.

The increase of Mg concentration eventually leads to the formation of bulk $XMg_2$ structures or coexistence of $XMg_2$ and X structures. The diffusion behavior of Mg atoms can then be described as diffusion in defected bulk $XMg_2$ structures. Ionic conductivity of such compound strongly depends on Mg vacancy concentration. In the simplest case, for an almost fully charged anode, it can be analyzed as the diffusion of a lone Mg vacancy (Mg atom) between two tetragonal positions (see Fig. S1) in $XMg_2$ structures. This still provides a general understanding of diffusion properties in prospective anode materials at high states of charge. For systems with a single Mg vacancy in an $XMg_2$ structure, three different migration pathways exist (see Fig. 5). Similar to investigations of oxygen diffusion in cubic zirconia [27] and ceria [28], we find that the lowest migration barrier corresponds to Mg migration along the [100] direction. For other migration pathways, cation-cation interactions significantly increase (by more than 0.5 eV) the migration barriers (see Tab. 2). Mg diffusion in $XMg_2$ structures can thus be described as motion of a Mg atom between two nearest Mg sites and the contributions from other pathways can be ignored. These results suggest that for the $XMg_2$ with the same concentrations of Mg vacancies, Ge based materials should have the highest ionic conductivity among the considered $XMg_2$ structures.

Since the charging/discharging rate is more important than the volumetric energy density for bulk storage applications [29], the trend in diffusion barriers among the 3 materials at different Mg concentrations computed here is of practical significance for electrode design. For instance, the low ionic conductivity of Si suggests that its use as an anode material for bulk storage Mg-ion batteries is questionable. It is also expected that due to the possible existence of inhomogeneous structures during metal insertion and high Mg diffusion barriers in Mg-poor structures, it may be difficult to achieve the maximum theoretic volumetric energy density for Si based materials. In contrast, both Ge and Sn, having low Mg diffusion barriers (comparable with migration barrier of Li atoms in Si (0.61 eV) [26]) for a range of dopant concentrations, are attractive anode materials for both high volumetric energy density and bulk storage applications.

**Conclusions**

In summary, based on DFT calculations, we have investigated the potential of α-X (X=Ge, Si, and Sn) systems as anode materials for Mg batteries. We find that all considered materials can provide comparable capacities and larger volumetric energy densities compared to those for Li batteries. The results show that Sn could provide the largest volumetric energy density (7.413 Wh cc$^{-1}$) at a 100% volume expansion and the lowest migration barrier for the diffusion of a single Mg atom (0.497 eV). Hence, using stabilized diamond Sn can provide the best performance among the considered materials. A slightly weaker performance can be observed for the Ge anode, which can provide a comparable volumetric energy density (7.382 Wh cc$^{-1}$) and slightly larger migration barriers for low dopant concentrations. Hence, Ge and Sn based materials can be attractive anode materials for both high volumetric energy density and bulk storage applications. As the concentration of Mg atoms increases, Mg-Mg interactions are expected to drive down the diffusion barrier by as much as 0.38 eV, a similar effect to that predicted by us for Li and Na diffusion in Si [26]. The barrier in the XMg$_2$ phase is also much lower than that at the beginning of the charging process. Therefore, while most theoretical studies of diffusion in battery electrode materials were done for a single dopant atom or ion [30, 31], we clearly demonstrate that the inclusion of metal-metal interactions is required for a realistic computational analysis of electrode materials.

**Supporting information**

Details on methods, structures, and defect formation energies are available as supporting information.

**Acknowledgments**

This work was supported by the Tier 1 AcRF Grant by the Ministry of Education of Singapore (R-265-000-430-133). We gratefully acknowledge the use of supercomputers in A-STAR Computational Resource Centre (ACRC) and support from the Institute of High Performance Computing of A-STAR Singapore.

**REFERENCES**

[1] V. Palomares, P. Serras, I. Villaluenga, K.B. Hueso, J. Carretero-Gonzalez, T. Rojo, Energy Environ. Sci., 5 (2012) 5884-5901.

[2] V.L. Chevrier, G. Ceder, J. Electrochem. Soc., 158 (2011) A1011-A1014.

[3] S.W. Kim, D.H. Seo, X.H. Ma, G. Ceder, K. Kang, Adv. Energy Mater., 2 (2012) 710-721.

[4] T.S. Arthur, N. Singh, M. Matsui, Electrochem. Commun., 16 (2012) 103-106.

[5] N. Singh, T.S. Arthur, C. Ling, M. Matsui, F. Mizuno, Chem. Commun., 49 (2013) 149-151.

[6] S. Liu, J.J. Hu, N.F. Yan, G.L. Pan, G.R. Li, X.P. Gao, Energy Environ. Sci., 5 (2012) 9743-9746.

[7] M. Hayashi, H. Arai, H. Ohtsuka, Y. Sakurai, J. Power Sources, 119–121 (2003) 617-620.

[8] D. Aurbach, Z. Lu, A. Schechter, Y. Gofer, H. Gizbar, R. Turgeman, Y. Cohen, M. Moshkovich, E. Levi, Nature, 407 (2000) 724-727.

[9] D. Aurbach, I. Weissman, Y. Gofer, E. Levi, Chem. Rec., 3 (2003) 61-73.

[10] J.O. Besenhard, M. Winter, ChemPhysChem, 3 (2002) 155-159.

[11] T. Ichitsubo, T. Adachi, S. Yagi, T. Doi, J. Mater. Chem., 21 (2011) 11764-11772.

[12] R. Zhang, X. Yu, K.-W. Nam, C. Ling, T.S. Arthur, W. Song, A.M. Knapp, S.N. Ehrlich, X.-Q. Yang, M. Matsui, Electrochem. Commun., 23 (2012) 110-113.

[13] M.N. Obrovac, L. Christensen, D.B. Le, J.R. Dahnb, J. Electrochem. Soc., 154 (2007) A849-A855.

[14] I.-H. Jung, J. Kim, J. Alloys Compd., 494 (2010) 137-147.

[15] M.K. Aydinol, A.F. Kohan, G. Ceder, K. Cho, J. Joannopoulos, Phys. Rev. B, 56 (1997) 1354-1365.

[16] M. Mortazavi, J. Deng, V.B. Shenoy, N.V. Medhekar, J. Power Sources, 225 (2013) 207-214.

[17] H. Wu, Y. Cui, Nano Today, 7 (2012) 414-429.

[18] M.Y. Ge, J.P. Rong, X. Fang, C.W. Zhou, Nano Lett., 12 (2012) 2318-2323.

[19] H. Wu, G. Zheng, N. Liu, T.J. Carney, Y. Yang, Y. Cui, Nano Lett., 12 (2012) 904-909.

[20] N. Liu, H. Wu, M.T. McDowell, Y. Yao, C. Wang, Y. Cui, Nano Lett., 12 (2012) 3315-3321.

[21] T.T. Tran, M.N. Obrovac, J. Electrochem. Soc., 158 (2011) A1411-A1416.

[22] E.A. Fitzgerald, P.E. Freeland, M.T. Asom, W.P. Lowe, R.A. Macharrie, B.E. Weir, A.R. Kortan, F.A. Thiel, Y.H. Xie, A.M. Sergent, S.L. Cooper, G.A. Thomas, L.C. Kimerling, J. Electron. Mater., 20 (1991) 489-501.

[23] J.Z. Hu, I.L. Spain, Solid State Commun., 51 (1984) 263-266.

[24] N.E. Christensen, M. Methfessel, Phys. Rev. B, 48 (1993) 5797-5807.

[25] W.H. Wan, Q.F. Zhang, Y. Cui, E.G. Wang, J. Phys.: Condens. Matter., 22 (2010) 415501.

[26] O.I. Malyi, T.L. Tan, S. Manzhos, submitted to Applied Physics Express.

[27] O.I. Malyi, P. Wu, V.V. Kulish, K. Bai, Z. Chen, Solid State Ionics, 212 (2012) 117-122.

[28] M. Nakayama, M. Martin, Phys. Chem. Chem. Phys., 11 (2009) 3241-3249.

[29] E. Barbour, I.A.G. Wilson, I.G. Bryden, P.G. McGregor, P.A. Mulheran, P.J. Hall, Energy Environ. Sci., 5 (2012) 5425-5436.

[30] Q.F. Zhang, W.X. Zhang, W.H. Wan, Y. Cui, E.G. Wang, Nano Lett., 10 (2010) 3243-3249.

[31] S.C. Jung, Y.-K. Han, Phys. Chem. Chem. Phys., 13 (2011) 21282-21287.

**Figure Captions**

**Figure 1.** Volumetric energy density for Mg-X alloys as a function of volume expansion.

**Figure 2.** Tetragonal (a) and hexagonal (b) insertion sites in Ge, Si, and Sn (=X). Orange – Mg atoms, blue – X atoms.

**Figure 3.** Formation energies of Mg-Mg defects as functions of Mg-Mg distance in Ge (a), Si (b), and Sn (c).

**Figure 4.** (a) Migration barriers for diffusion of Mg atom between two tetragonal sites of Ge, Si, and Sn. (b) The migration barriers of Mg in bulk Ge, Si, and Sn vs. Mg-Mg distance in the bulk materials.

**Figure 5.** Possible migration pathways for a Mg (vacancy) atom in $XMg_2$ structures (X=Ge, Si, Sn). Red, green, and grey colors represent [100], [110], and [110] directions for vacancy migrations. Orange – Mg atoms, grey – Mg vacancy, blue – X atoms

TABLES

Table 1 Specific capacity (mAh g$^{-1}$), percent volume expansion (%) at full state of charge, and voltage (V) for $GeMg_2$, $SiMg_2$, and $SnMg_2$

| Element | Specific capacity | Volume Expansion | Voltage |
|---------|-------------------|------------------|---------|
| Ge | 1476 | 178 | 0.241 |
| Si | 3817 | 216 | 0.151 |
| Sn | 911 | 120 | 0.184 |

Table 2 Migration barriers (in eV) of single Mg vacancy in $XMg_2$ structures

| | [100] | [110] | [111] |
|---|---|---|---|
| $GeMg_2$ | 0.419 | 1.378 | 1.418 |
| $SiMg_2$ | 0.444 | 1.438 | 1.475 |
| $SnMg_2$ | 0.497 | 0.948 | 0.955 |

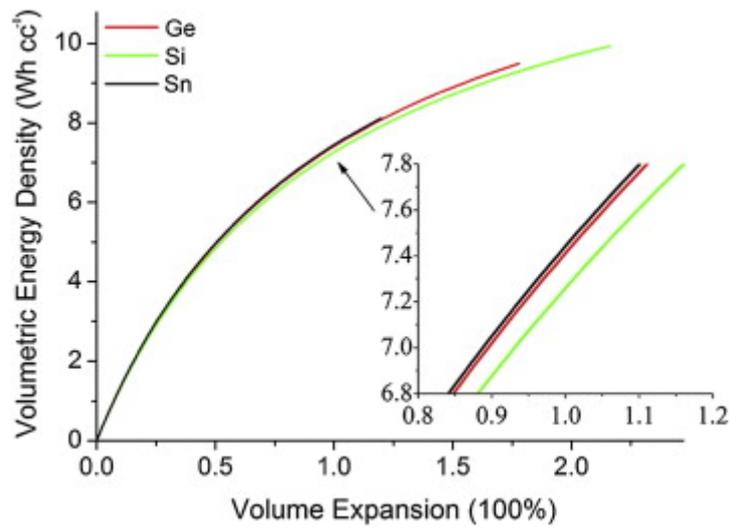

**Figure 1**

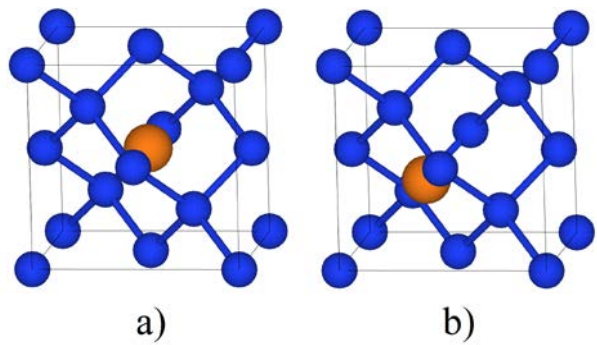

**Figure 2**

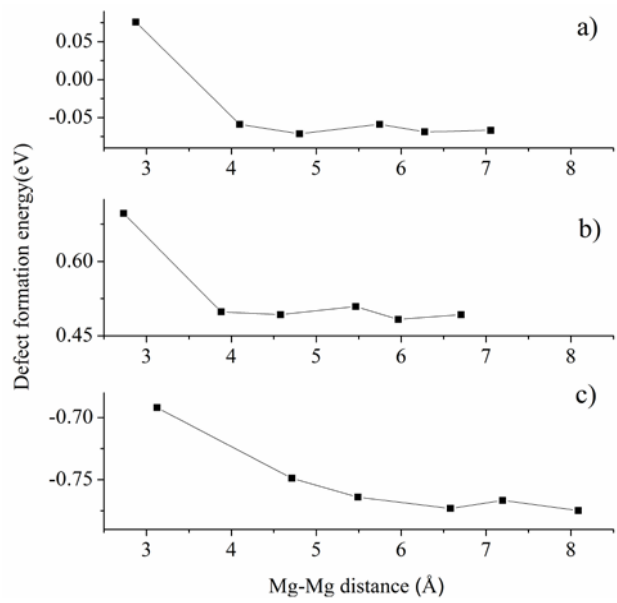

**Figure 3**

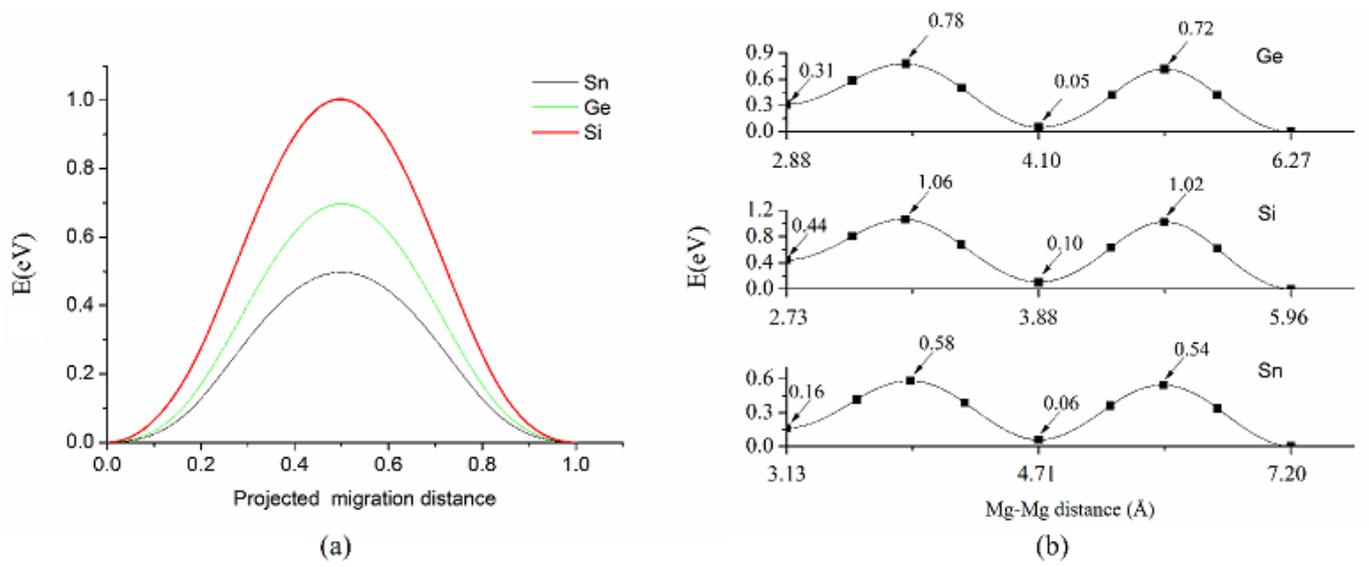

**Figure 4**

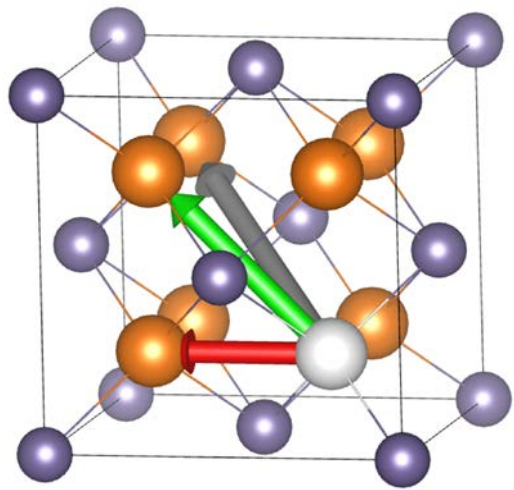

**Figure 5**